\documentclass{mem}
\usepackage{natbib}\usepackage{txfonts}\usepackage{balance}
\usepackage{flushend}
\usepackage{graphicx}
\usepackage[breaklinks,pdftex]{hyperref}
\idline{75}{282}
\begin{document}
\def\teff{$T\rm_{eff }$}
\def\kms{$\mathrm {km s}^{-1}$}

\title{
Recent advances in the photometric investigation of classical pulsators with the \textit{Kepler}/K2 and TESS missions
}

\author{
E. \,Plachy\inst{1,2,3} 
          }

\institute{
Konkoly Observatory, Research Centre for Astronomy and Earth Sciences, MTA Centre of Excellence, Konkoly Thege Mikl\'os \'ut 15-17, H-1121 Budapest, Hungary
\and
MTA CSFK Lend\"ulet Near-Field Cosmology Research Group
\and 
ELTE E\"otv\"os Loránd University, Institute of Physics, 1117, P\'azm\'any P\'eter s\'et\'any 1/A, Budapest, Hungary\\
\email{eplachy@konkoly.hu}
}

\authorrunning{Plachy}

\titlerunning{Space photometry}

\date{Received: Day Month Year; Accepted: Day Month Year}

\abstract{
Nearly continuous, densely sampled, space-based photometry allows us to recover the finest details in the light variations of stars. The number of such light curves have been increasing rapidly in the last few years thanks to the extended mission of the \textit{Kepler} space telescope and the launch of the TESS mission. This new era brings us new perspectives in RR~Lyrae and Cepheid studies, where low amplitude phenomena can be studied in a wide range of individual stars and on a statistical basis. In this proceedings I review the recent investigations of the \textit{Kepler}, K2 and TESS fields, as well as the challenges in accurately reducing high-quality photometry. 

\keywords{Stars: Variables: RR~Lyrae stars, Cepheids -- Techniques: Space photometry}
}
\maketitle{}

\section{Introduction of the \textit{Kepler}, K2 and TESS missions}

The \textit{Kepler} mission targeted a 105 square degree area near the Cygnus and the Lyra constellations \citep{borucki-2010}. Forty-one RR~Lyrae stars and two Cepheids were observed in this field. The number of observed classical pulsators was low, but the precision and the coverage of the photometry was exceptional, which led us to new types of investigations and many discoveries in the millimagnitude range of the light variations.

The telescope lost its pointing ability after 4 years of operation due to a gyroscope failure. The \textit{Kepler} mission ended with this incident, but a new observing mode was conceived and was named the K2~mission. In this new mission different fields were observed for $\sim$80 days along the ecliptic plane \citep{howell-2014}. In these so called Campaigns more 3000 RR~Lyrae stars and few dozens of Cepheids were proposed to be observed in the Guest Observer Program. These targets were collected from large variability catalogs available at that time (see references in \citealt{plachy-2016}). Even more RR~Lyrae and Cepheids can be found in the superstamps defined towards dense stellar fields in the galactic bulge and globular clusters.
These spatially large targets were observed using many smaller stamps that can be tiled together to form a complete superstamp image.

The TESS (Transiting Exoplanet Survey Satellite) space telescope started observing in July of 2018, and a few month later the K2~mission ended when fuel run out in Campaign~19. TESS is a near all-sky mission, which divides the sky into overlapping sectors that are observed for 27 days, respectively \citep{ricker-2015}. A fundamental difference between \textit{Kepler} and TESS observation strategies is that in \textit{Kepler} and K2 only pre-selected pixels masks were downloaded from the spacecraft, while with TESS it is possible to frequently download the full frame images (FFI) with 30-minute cadence. That provides us with light curves for all stars on the FFIs within the 16 magnitude brightness limit, including several tens of thousands of RR~Lyrae stars and most Galactic Cepheids. In the first year of the mission, TESS observed the southern ecliptic hemisphere with a continuous viewing zone around the Southern Ecliptic Pole. In the second year, the northern ecliptic hemisphere was observed, but a large area has been skipped because of the presence of scattered light from the Earth and the Moon. TESS has a very stable orbit, thus we can expect a series of extended missions, in which the full frame images can be taken even more frequently. In Extended Mission 1 the cadence was elevated to 10 minutes, and will be increased further in the second extension to 200 seconds. In the extended missions the ecliptic plane is also being observed. 

Both \textit{Kepler} and TESS have broad passband filters, \textit{Kepler} is centered on the Cousins \textit{R }band,  TESS is on the \textit{I} band. Therefore TESS light curves are comparable with \textit{I}-band data of the Optical Gravitational Lensing Experiment (OGLE \citealt{ogle-1992}) and Gaia $G_{\rm RP}$-band measurements, and \textit{Kepler} is compatible with Gaia \textit{G}-band \citep{gaia-bands}.
Regarding the angular resolution there is a large difference between \textit{Kepler} and TESS: they have 4 and 21 arcsecond per pixel resolutions, respectively. The low spatial resolution of TESS is the source of possible blending and contamination from neighboring stars.  

\section{Photometric challenges}

With space telescopes we avoid the effects of the atmosphere, but we have other issues to deal with instead. A great challenge has been for example the combination of the data from different Quarters of the \textit{Kepler} mission. Variation due to the 90-degree roll of the telescope in every 3 months, and the change in thermal properties of the telescope elements during the year strongly affected the light curves. The instrumental signals could be eliminated only with proper detrending and scaling before the data stitching \citep{kepler-stitching}.

The precision of the pointing was degraded in the K2 mission and therefore correction maneuvers had to be taken at every six hours. This caused changes in the position of the stars on the detector, which periodicity could appear in the Fourier spectra of the light curves. Several pipelines have been developed to correct the systematics. However, the similarity between the shapes of the systematics and the RR~Lyrae variation deceives correction pipelines in many cases. Some methods do not handle well the fast and large amplitude variations either. After recognizing this problem, a new photometric pipeline has been invented that is optimized for RR~Lyrae type variation.   

The pipeline uses apertures extended to contain the PSF of the star at any time during the observations \citep{eap}. This extension already reduces the systematics significantly that can be corrected even further. The pipeline gives equally good or better solutions compared to other pipelines for the majority of RRL stars, sometimes providing the only useful light curve.
Nearly 3000 RR~Lyrae light curves have been processed and published by \citet{autoeap} along the code, in which the pixel selection process is automatized and can be used for other types of variable stars as well. 

The main photometric challenge of the TESS mission was expected to be contamination from neighboring stars due to the large pixel size of the detectors, but then it turned out that the scattered light is a larger problem. TESS operates in a highly eccentric 2:1 lunar resonance orbit. At certain times, the light from the Earth and the Moon can be seen by the cameras directly or is reflected by the lens hood. This causes either an overall or a localized brightness increase in the sky background, sometimes to the point of saturation \citep{handbook}. The light curves of targets stars on contaminated pixels therefore might be partially or completely useless. 

Each sector consists of two consecutive orbits. The data downlink occurs in every orbit, when the spacecraft is reoriented to point the transmitter at the Earth, and data collection is interrupted. Therefore data gaps are present not only between the sectors but at the middle of the sectors as well, leading us to a similar data stitching problem that was experienced in the original \textit{Kepler} mission. 

The TESS mission provides light curves only for pre-selected high cadence targets, among which only few RR~Lyrae and Cepheids can be found. The FFIs are open to the scientific community to prepare the light curves for themselves. Data releases that includes RR~Lyrae and Cepheid light curves have already been published. The Quick Look Pipeline provides light curves for all observed stars within T$<$13.5 mag brightness limit in TESS magnitude for the primary and the first extended missions \citep{qlp-0,qlp-1, qlp-2}. The TESS Data for Asteroseismology (T'DA) group of the TESS Asteroseismic Science Consortium (TASC) has produced light curves for all stars down to a TESS magnitude of 15 \citep{handberg-2021,lund-2021}. At the time of writing this paper the T'DA light curves are accessible for Sectors 1-6, 14-15 and 26.  The pipeline by \citet{powell-2022} computes light curves for all stars in the FFIs of the primary mission brighter than 16 magnitudes. All TESS data products mentioned above are publicly available at the Mikulski Archive for Space Telescopes (MAST\footnote{\url{https://archive.stsci.edu/}})

Custom apertures for individual TESS targets can be prepared with the open-source \texttt{Lightkurve} tool \citep{lightkurve}, while the popular \texttt{eleanor} tool provides different photometric solutions, including PSF photometry.

\section{Pulsation seen by the \textit{Kepler} mission}

We have learned with from the space-based missions that a noise level of a millimagnitude can be reached in the amplitude spectra, thus allowing us to reveal new phenomena in the pulsation of RR~Lyrae and Cepheid stars. In the \textit{Kepler} mission four main topics has been studied with unprecedented accuracy. (1) The low amplitude additional mode are easily detectable in the residual Fourier spectra of the \textit{Kepler} data, after the main pulsation has been removed. (2) The discovery of period doubling has been possible due to the high cadence and continuity of light curves. (3) The fine details of the Blazhko effect \citep{blazhko-1907} could be monitored over four years. (4) The same is true for the small scale instabilities in the O-C diagrams. These types of investigations are continuing with K2 and TESS. 
For a detailed review of the \textit{Kepler} results on RR~Lyrae stars we refer to \citet{frontiers}. The only classical Cepheid in the \textit{Kepler} field (V1154~Cyg) has been studied by \citet{derekas-2012,derekas-2017,kanev-2015}, while the only type~II Cepheid (DF Cyg) has been analysed by \citet{dfcyg-bodi,dfcyg-vega,dfcyg-plachy}.

Many years after the end of the original \textit{Kepler} mission, it still provides us with data to investigate. During the \textit{Kepler} mission almost all RR~Lyrae targets were measured in 1-minute cadence mode at least for one quarter length. These data were studied, and cycle-to-cycle variations in the order of a few millimagnitudes were revealed \citep{benko-2019}.
The long cadence data were also re-analysed with time-dependent methods and showed that the additional low-amplitude modes can be excited either permanently or temporarily.

The systematic search in the background pixels of the original \textit{Kepler} field (the \textit{Kepler} Pixel Project) led to the detection of 26 additional RR~Lyrae stars \citep{forro-2022}. Five of them are entirely new discoveries. The precision achievable from the background pixels is adequate to study Blazhko effect, too.

\section{Investigations of the K2 fields}

RR~Lyrae candidates have been proposed for observation in each K2 Campaign, from which altogether about 4000 has been observed. Fifteen percent of these stars turned out to be misclassified variables, while for another 20 percent of them no good-quality light curves are available.

Nearly 2600 K2 RR~Lyrae stars can be collected for analysis from different photometric pipelines, but the majority of the best quality light curves are from the Extended Aperture Photometry. Eighty-one percent of the sample are RRab stars, $\sim$17 percent are RRc, and $\sim$2 percent are RRd. These incidence rates, however, are strongly biased by selection effects, as targets had to be identified from existing surveys.

\begin{figure*}[t!]
\resizebox{\hsize}{!}{\includegraphics[clip=true]{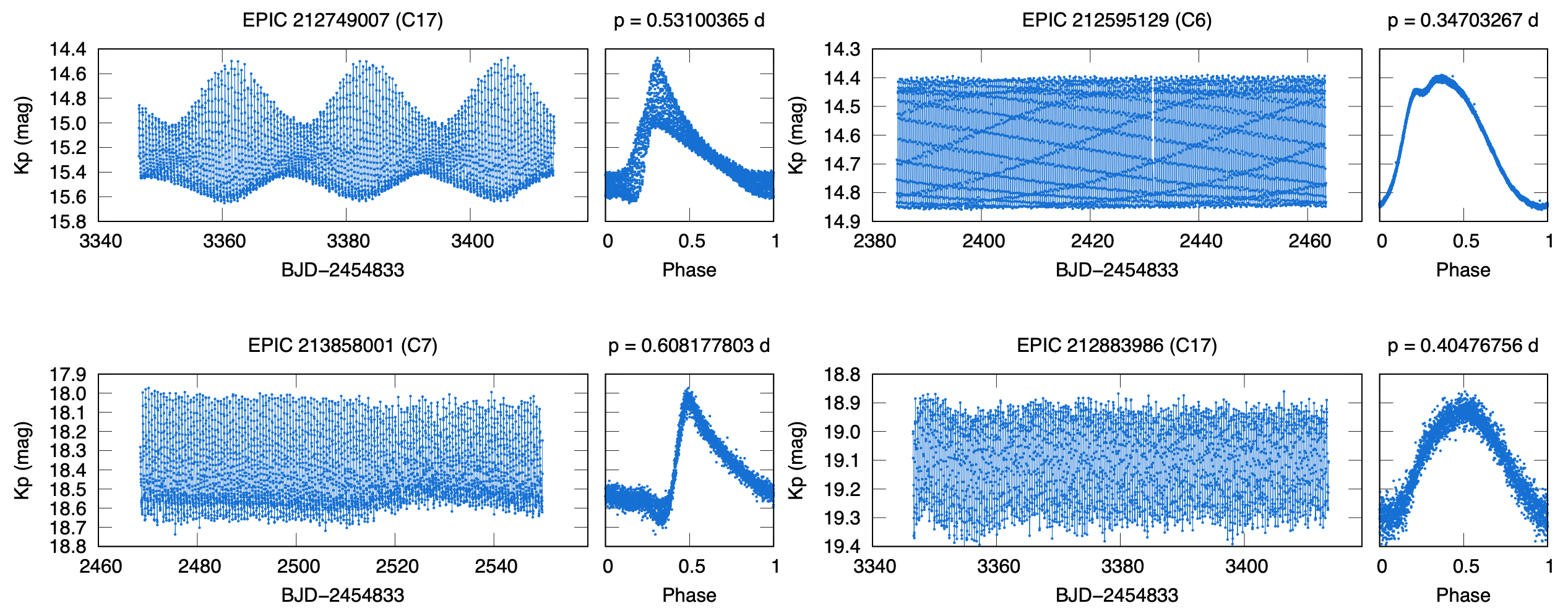}}
\caption{\footnotesize
{ Example light curves of K2 RR Lyrae stars published by \citet{autoeap}. }
}
\label{fig:rrl}
\end{figure*}

Most of the light curves are precise enough to identify additional low-amplitude modes and the Blazhko modulation. The latter is not possible if the modulation occurs at timescales much longer than the length of the data (80 days). A small group of light curves remained low-quality even after all the correction efforts that has been made. From those data we cannot detect modulation or additional modes reliably. Example light curves are presented in Fig.~\ref{fig:rrl}. The upper panels show high-quality K2 light curves, while the lower panels show data affected by instrumental effects, which is common for faint stars in dense stellar fields.    

\subsection{Search for the Blazhko effect}

The Blazhko effect manifests itself in the form of side peaks in the Fourier spectrum near the main pulsation frequency and its harmonics. The search of the side peaks is a widely used method to detect the Blazhko effect. However, in the K2 data side peaks can be seen due to the instrumental effects, too, and may remain even after light curves corrections have been applied. On the other hand, the continuity of K2 light curves allows us to study the change of amplitude and a phase synchronously. This was used in the study of \citet{eap} to determine the occurrence rate of the Blazhko effect in the early K2 Campaigns.
The relation of the amplitude and phase changes are being investigated further in the full K2 RR~Lyrae sample (B\'odi et al., in prep).

\subsection{Additional modes}

The Petersen diagram \citep{petersen} provides us with a spectacular representation of the period relations of multimode pulsation. Seventy-seven RRd stars have recently been analysed in the K2 RR~Lyrae sample \citep{moskalik-2021}. Three of these stars are possible members of the anomalous RRd group of which the period ratios fall outside of the classical ones. Four other RRd stars are marked as `peculiar' \citep{nemec-2021}, showing resemblance to the distinct short period group discovered by \citep{prudil-2017}. Three of the K2 `peculiar' RRds were already known, one is a new discovery which is bright enough for us to study the temporal variation of the modes. Both modes has been found to be nonstationary. 

The low-amplitude additional modes are also being studied in the Petersen diagram, where different groups are seen to be forming at 0.61 and 0.68 period ratios with the radial overtone mode (the former have shorter while the latter have longer periods than the overtone pulsation). The main source of discoveries in low-amplitude modes in RRc stars has been the OGLE Survey (see \citealt{smolec-2017} and \citealt{netzel-2019} and references therein).  For the first time, massive analysis of the low-amplitude modes can be performed with space photometric data as well. The K2 RR~Lyrae sample includes $\sim$450 RRc stars, and the results suggest that low-amplitude modes are present over 60 percent of them \citep{netzel-2023}. 

The fundamental-mode RR~Lyrae stars also exhibit low-amplitude modes, among which the second radial overtone has already been identified \citep{molnar-2017}. Another group can be detected at $\sim$0.66-0.71 period ratios with the fundamental mode. The 0.66 ratio corresponds to the $1.5F_0$ frequency peak caused by the period doubling phenomenon \citep{kolenberg-2010}. It is not clear yet whether the frequencies shifted to higher period ratios could also be the sign of the period doubling.

\subsection{Globular clusters}

Several globular clusters have been observed in the K2 mission, but only few of them have been analyzed already. M4 is the closest one, where sub-millimagnitude precision could be achieved in the photometry. Two interesting variables were found here, just outside the RR~Lyrae instalibility strip, so they are proposed to be millimagnitude RR~Lyrae candidates \citep{wallace-2019}. 

In M80 we know more overtone RR~Lyrae stars than fundamental mode ones, which makes this globular cluster very interesting. Some of the RRc stars show peculiar modulations that are multiperiodic or irregular. K2 photometry of M80 RR~Lyrae stars have high enough quality to detect low-amplitude additional modes as well (Moln\'ar et al. in prep). 

\subsection{K2 Cepheids}

Few dozen Cepheid stars were observed in the K2 mission. A recent study of type~II and anomalous Cepheids \citep{monika-2022} revealed low amplitude features among them: cycle-to-cycle variations at millimagnitude level were detected in all analysed W~Vir stars. The short-period BL Her and anomalous Cepheids were also found to be more stable in the O-C diagrams than the longer-period ones. 

\section{First results from the TESS mission}

The TESS mission provides us with a unique opportunity to study low-amplitude features with space photometry from nearly the entire sky. But it was not clear before the mission which kinds of features can be studied with TESS and at what magnitude limit. Two studies has been published by the Cepheid and RR Lyrae Working Group of TASC to explore the potential of TESS data, one was aimed the RR~Lyrae \citep{tess-rrl} while the other the Cepheid stars \citep{tess-cep}.   

\subsection{RR~Lyrae stars with TESS}

The first investigations focused on a sample of 118 RR~Lyrae stars that were chosen from Sectors 1 and 2 \citep{tess-rrl}. It was shown that classification via the precise light curve shapes combined with parallax information from the Gaia space mission \citep{gaia-edr} is very effective, and short period anomalous Cepheid candidates can be separated. The one-sector-long light curves are generally too short for detailed studies of the Blazhko effect, and its occurrence rate can also be estimated only with high uncertainty. 

Nevertheless, TESS data is ideal to hunt for low-amplitude extra modes and other signals, and  various types of extraperiodicities were found of in a large fraction of stars. In the case of RRab stars they are grouped in three broad regions in the Petersen diagram that potentially correspond to the first and second overtone mode, and the period doubling, while the spread of period ratio values remains puzzling. The TESS data also confirmed the previous experience that additional modes in RRab stars always coincide with the presence of the Blazhko modulation. The incidence rates of 0.61 and 0.68 modes of RRc stars was found to be 65 and 16 percent, respectively, nearly an order of magnitude higher than in the OGLE Bulge sample \citep{netzel-2019}. RR~Lyrae stars with extra modes were found to be more frequent towards the center of the instability strip. It has been concluded that RRab-type variation is still recognizable for T $<$ 17~mag, but the Blazhko effect and the low amplitude modes are barely or not visible beyond 15.5 mag.

A massive analysis of RRc stars observed in the TESS primary mission suggest that low-amplitude modes are more frequent than previously found to be \citep{tess-rrc}.

In the huge RR Lyrae sample monitored by TESS we can find many stars that are interesting from a certain aspect. One such example was shown by \citealt{carrell-2021}, who presented a double-mode RR~ 
Lyrae star (V338~Boo) where both modes show temporal variations. Blazhko-type modulation is frequent phenomenon in anomalous RRd stars seen by OGLE \citep{smolec-2015} and was also investigated in K2 data (\citep{mod_rrd}). V338~Boo is the first Blazhko candidate among classical RRd stars.

\subsection{TESS Cepheids}

For RR~Lyrae stars we are confined to the Milky Way with TESS. However, when it comes to Cepheids, we are able to observe extragalatic ones with it, too \citep{tess-cep}. (1) An 0.61 mode was found in an overtone Cepheid in the Large Magellanic Cloud beyond the detection limit of OGLE this way. (2) Beside the extragalactic objects, different types of Cepheid were investigated within the Galaxy, too, and the 0.61 mode was detected for the first time in an anomalous Cepheid, XZ Cet. (3) The sign of Blazhko type modulation was seen in another overtone anomalous Cepheid, AK PsA. (4) A special object, $\beta$ Dor, was also investigated in the continuous viewing zone of TESS. $\beta$ Dor was observed in 2-minute cadence mode, but due to its brightness (T=3.26 mag) it was saturated in the TESS images. Custom aperture photometry was applied to cover the entire saturation column both at mimimum and maximum brightness. This improved light curve showed intrinsic cycle-to-cycle variations in the order of  5-10 millimimagnitudes. 

\subsection{TESS photometry as auxiliary data}

\begin{figure*}[t!]
\resizebox{\hsize}{!}{\includegraphics[clip=true]{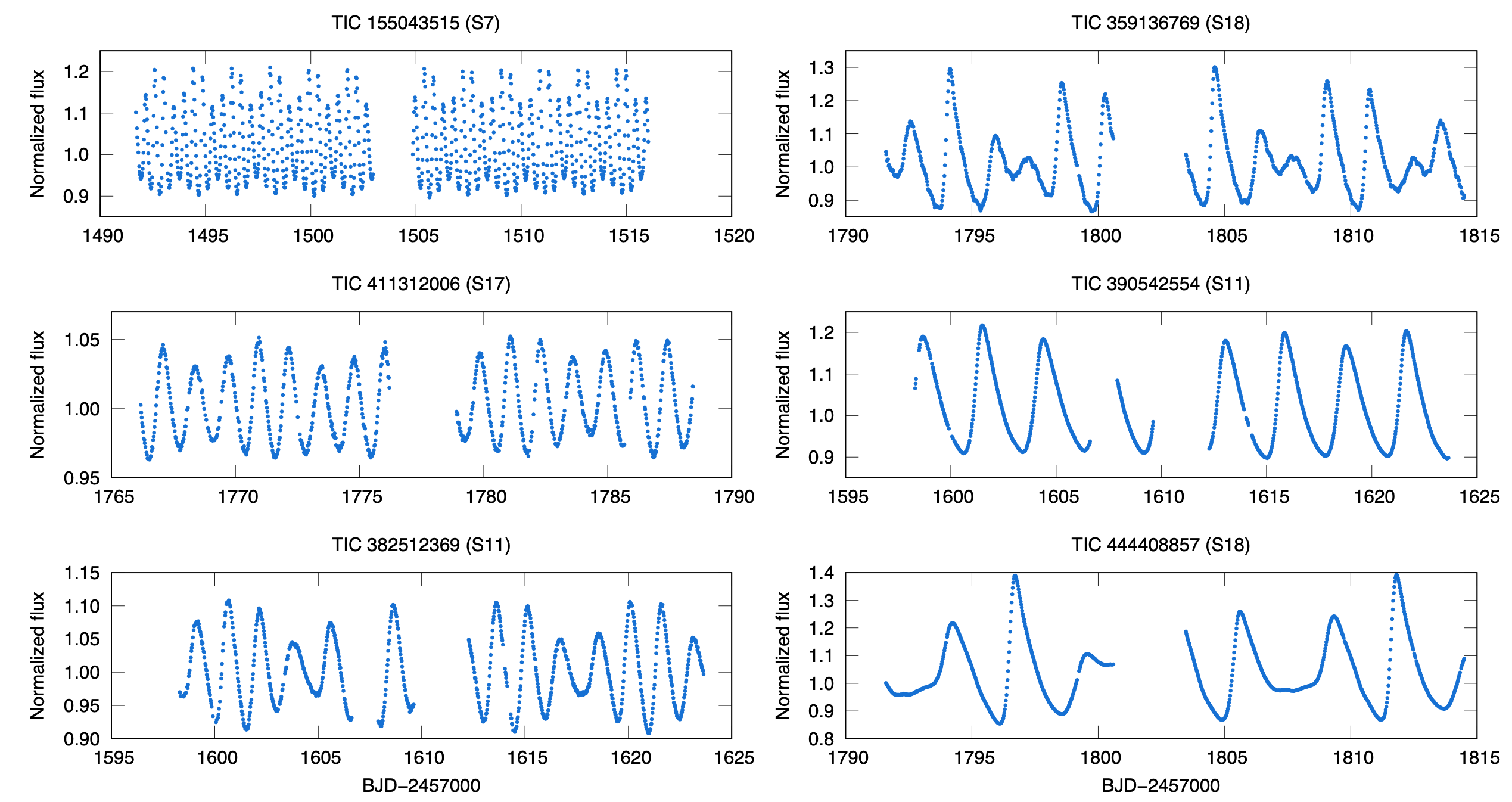}}
\caption{\footnotesize
{Example light curves of TESS double-mode Cepheids from the QLP photometric pipeline published by \citet{qlp-1}. }
}
\label{fig:cep}
\end{figure*}

TESS observes many of the RR~Lyrae and Cepheid stars that are accessible with ground-based spectroscopy, thus it is straightforward to use TESS light curves as auxiliary information in spectroscopic studies. A main strength of TESS data is that it can reveal secondary modes even at at very low brightness level \citep{ripepi-2021,benko-2021}. The same conclusion was made during validation work of the \textit{Gaia} DR3 Cepheid candidates \citep{ripepi-2022}, where significant number of further double-mode stars cloud be classified with TESS compared to the \textit{Gaia} identifications. Few examples of double-mode Cepheid light curves are presented in Fig.~\ref{fig:cep}.

\section{Conclusions}

As this brief review illustrates, the \textit{Kepler} and TESS space telescopes provide us with a giant leap forward in photometric data, and through that, new insights into the low-amplitude features of the pulsation of RR~Lyrae and Cepheid stars. The \textit{Kepler} mission, although limited in the number of targets that were observed, still has unexplored potential, thanks to its multi-year continuous coverage of stars. 

Given the need for more intensive corrections in K2 mission, the data processing was delayed for these data sets. But we understand the instrumental characteristics much better now, and know how to correct them while keeping the pulsation signals intact. Massive analysis of RR~Lyraes and Cepheids observed by K2 has started, and will continue in the future. The mission covered thousands of RR~Lyrae stars, from the halo and the bulge to globular clusters and the Sagittarius stream, as well as several Cepheids, giving us much wider diversity of targets than the original mission could.

In the TESS mission the data stream is much faster than the speed of the photometric data processing, especially when it comes to pulsating stars. Nevertheless, FFI light curves of the primary and the first extended missions are already available from certain pipelines. The first larger studies based on TESS observations of RR Lyrae and Cepheid stars are now under way. The main strength of TESS is that it provides extremely valuable photometry from nearly the entire sky. It may have already acquired measurements for any brighter RR Lyrae or Cepheid star.

\begin{acknowledgements}
The research was supported by the LP2018-7 Lend\"ulet grant  of  the  Hungarian  Academy  of  Sciences, the  ‘SeismoLab’ KKP-137523 \'Elvonal grant of the Hungarian Research, Development  and  Innovation  Office  (NKFIH). The suggestions of the anonymous referee are gratefully acknowledged. \end{acknowledgements}

\bibliographystyle{aa}
\bibliography{bibliography}

\end{document}